\documentclass[journal]{IEEEtran}
\IEEEoverridecommandlockouts

\usepackage{amssymb}
\usepackage[cmex10]{amsmath}
\usepackage{stfloats}
\usepackage{graphicx}
\usepackage{subfigure}
\usepackage{tabularx}
\usepackage{verbatim}
\usepackage{url}
\usepackage{bm}
\usepackage{booktabs}
\usepackage[colorlinks]{hyperref}

\usepackage{algorithm}
\usepackage{algorithmic}

\usepackage{color}
\definecolor{myc1}{rgb}{0,0,0}

\begin{document}

\title{\LARGE Joint Beamforming and Antenna Design for\\Near-Field Fluid Antenna System}

\author{Yixuan Chen, 
            Mingzhe Chen, \IEEEmembership{Member, IEEE,}
            Hao Xu, \IEEEmembership{Member, IEEE,} 
            Zhaohui Yang, \IEEEmembership{Member, IEEE,}\\
            Kai-Kit Wong, \IEEEmembership{Fellow, IEEE,}
           and Zhaoyang Zhang, \IEEEmembership{Senior Member, IEEE}
\vspace{-8mm}

\thanks{This work was supported by the National Key R\&D Program of China (Grant No. 2023YFB2904804), National Natural Science Foundation of China (NSFC) under Grants 62394292, 62394290.}
\thanks{The authors thank the Technology Innovation and Training Center, Polytechnic Institute, Zhejiang University, Hangzhou, Zhejiang Province, China for providing the Millimeter-wave sensing interconnection hardware and software integrated system.}
\thanks{Y. Chen, Z. Yang, and Z. Zhang are with the College of Information Science and Electronic Engineering, Zhejiang University, and also with Zhejiang Provincial Key Laboratory of Info. Proc., Commun. \& Netw. (IPCAN), Hangzhou, 310027, China (e-mails: \{chen\_yixuan, yang\_zhaohui, ning\_ming\}@zju.edu.cn).}
\thanks{Mingzhe Chen is with the Department of Electrical and Computer Engineering, Institute for Data Science and Computing, University of Miami, Coral Gables, FL 33146 USA (e-mail: mingzhe.chen@miami.edu).}
\thanks{Hao Xu is with the Department of Electronic and Electrical Engineering, University College London, WC1E 7JE London, U.K. (e-mail: hao.xu@ucl.ac.uk).}
\thanks{Kai-Kit Wong is with the Department of Electronic and Electrical Engineering, University College London, WC1E 7JE London, U.K., and also with the Yonsei Frontier Laboratory, Yonsei University, Seoul 03722, South Korea (e-mail: kai-kit.wong@ucl.ac.uk).}
}

\maketitle

\begin{abstract}
In this letter, we study the energy efficiency maximization problem for a fluid antenna system (FAS) in near field communications. Specifically, we consider a point-to-point near-field system where the base station (BS) transmitter has multiple fixed-position antennas and the user receives the signals with multiple fluid antennas. Our objective is to jointly optimize the transmit beamforming of the BS and the fluid antenna positions at the user for maximizing the  energy efficiency. Our scheme is based on an alternating optimization algorithm that iteratively solves the beamforming and antenna position subproblems. Our simulation results validate the performance improvement of the proposed algorithm and confirm the effectiveness of FAS.
\end{abstract}

\begin{IEEEkeywords}
Near-field communication, fluid antenna system, statistical channel state information, energy efficiency.
\end{IEEEkeywords}
\IEEEpeerreviewmaketitle

\section{Introduction}
\IEEEPARstart{O}{ver the past} few decades, multiple-input multiple-output (MIMO) technologies have been integral for wireless communications systems. Traditionally, MIMO systems are based on employing fixed-position antennas at both sides and utilize signal processing to optimize the communication performance. While the capability of MIMO is undeniable and the extra-large (XL) version of MIMO is expected to become reality \cite{Wang-xlmimo}, it is doubtful if MIMO alone can match with the ever-growing demands in mobile communication systems. A new degree-of-freedom (dof) in the physical layer is needed.

To achieve this, fluid antenna system (FAS) and movable antenna (MA) have arisen as a new form of reconfigurable antenna technology for shape and position flexibility. MA utilizes mechanical structures to move flexibly in three-dimensional space\cite{li2022using}. Compared to traditional fixed-position antenna (FPA) systems, MA-assisted communication systems exhibit stronger spatial multiplexing performance and also demonstrate excellent interference resistance in beamforming \cite{10286328,10278220}.
FAS may be implemented using surface-wave technology \cite{Shen-tap_submit2024}, pixel-reconfigurable technology \cite{Zhang-pFAS2024} and others. It was first introduced to wireless communications by Wong {\em et al.}~in \cite{9131873,9264694}. In position-flexible FAS, the antenna has the ability to switch its position over a prescribed space. This has in recently years produced many interesting results, from performance analysis, to optimization and channel estimation, e.g., \cite{10103838,10318061,9992289,Vega-2023,Psomas-dec2023,10303274}. A recent article has provided a relatively updated list of literature \cite{LZhu24}.


With the increasing of operating frequency for more bandwidth, the good news is that FAS will be more powerful as its electrical size increases. Nevertheless, this also means that the conventional far-field plane wave assumption is no longer applicable since the base station (BS) will have an XL-MIMO array \cite{10220205}. Near-field communications is gathering increasing attention due to its importance, with recent attempt to clarify its difference from the far-field counterpart \cite{liu2023near} and methods addressing channel estimation in the near-field \cite{10078317}.

Motivated by the above discussion, it is important to optimize the BS and FAS at the user with consideration of near-field communications. In particular, in this letter, our aim is to jointly design the beamforming of the XL-MIMO at the BS and the antenna positions of the user's FAS provided the user has multiple fluid antennas, or a fluid antenna with multiple activated ports in the near field. Specifically, this letter has made the following contributions:
\begin{itemize}
\item We consider a downlink system, where a BS with multiple fixed-position antennas communicates to a user with FAS in the near field. The FAS at the user has multiple activated ports with variable positions. We formulate the maximization problem of system energy efficiency by jointly optimizing the antenna positions of FAS and the transmission covariance matrix at the BS.
\item To solve this energy efficiency maximization problem, we simplify the problem using an upper bound of the transmission rate. Subsequently, we decompose the original problem into two optimization subproblems, alternatingly optimizing the transmission covariance matrix and the antenna positions. An alternating optimization algorithm was proposed to find a solution to this problem.
\item Simulation results demonstrate the superiority of the proposed algorithm, showing that it achieves higher system energy efficiency compared to traditional schemes.
\end{itemize}
\section{System Model}
\begin{figure}[t]
\centering
\includegraphics[width=3.4in]{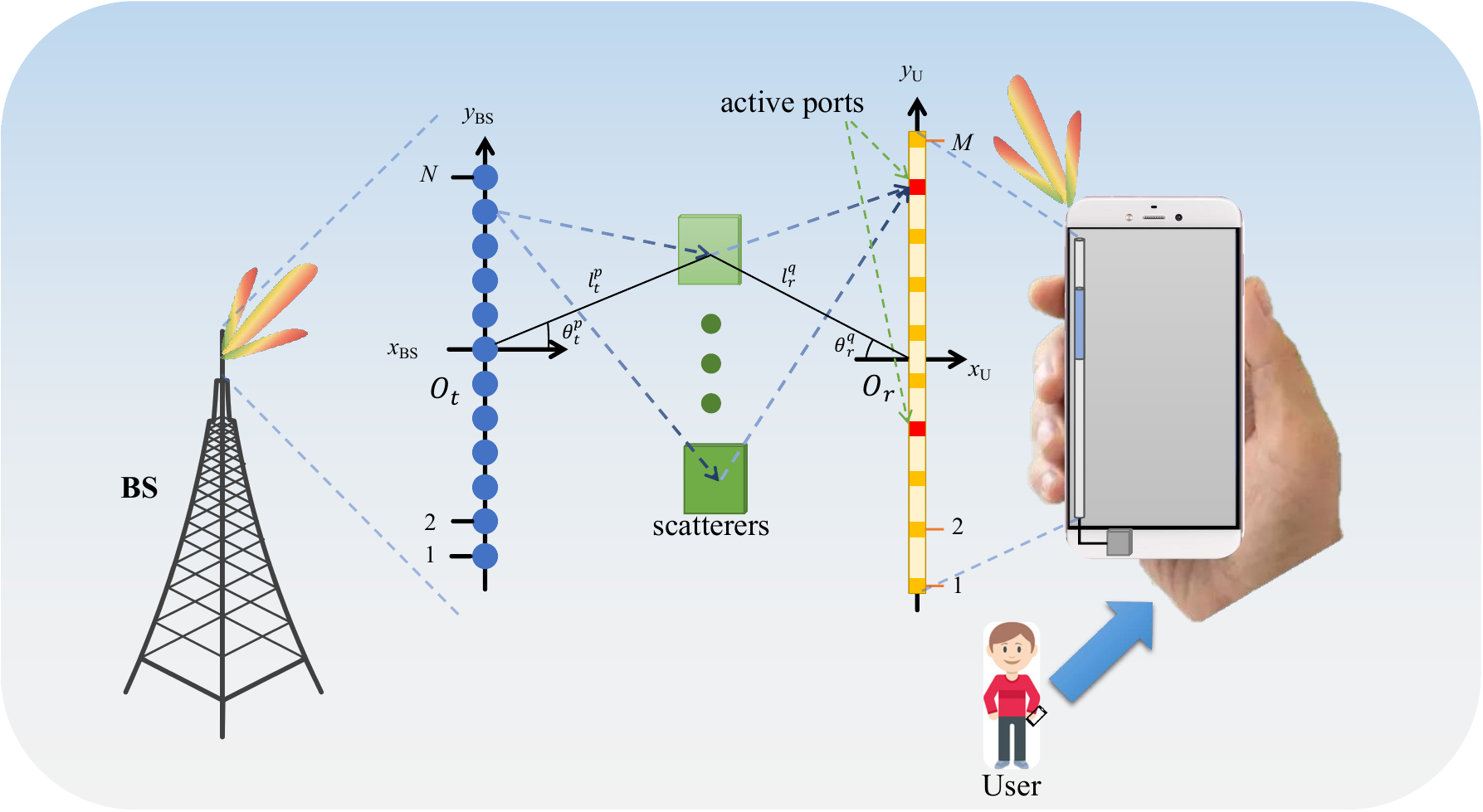}
\caption{The system model in the near-field with the coordinate systems.}\label{fig1}
\end{figure}

Consider a downlink communication system in the near field where a BS with multiple fixed-position antennas serves one user equipped with a fluid antenna that consists of $M$ ports (i.e., $M$ flexible positions, each of which can be activated if selected), as illustrated in Fig.~\ref{fig1}. Specifically, in this letter, we consider that more than one ports can be activated at any one time. Details will be discussed later in Section \ref{ssec:ap}. It is assumed that the channel has several scatterers (the subsequent model is not applicable to scenarios with few scatterers), and that the BS antenna size and carrier frequency are sufficiently large, the Rayleigh distance jointly determined by the BS and the user terminal will also be sufficiently large, allowing communication within the near-field range.
\subsection{Antenna Position Coordinates}\label{ssec:ap}
We establish a coordinate system with the center of the BS antenna array, $O_t$, as the origin. The transmit antennas of the BS are uniformly distributed along the $y_{\rm BS}$, perpendicular to the horizontal plane (along the $x_{\rm BS}$) with a spacing $d_{\rm BS}$. The total number of antennas at the BS is $N$. The coordinates of antenna $n$ are $\left(0,k_nd_{\rm BS}\right)$ where $k_nd_{\rm BS}$ is given as
\begin{equation}\label{eq1}
k_nd_{\rm BS}=\frac{2(n-1)-N+1}{2}d_{\rm BS}.
\end{equation}

We also establish a local Cartesian coordinate system with the center of the fluid antenna at the user, $O_r$, as the origin. The fluid antenna has its ports located along with the $y_{\rm U}$. The user terminal can have any posture so the $y_{\rm U}$ at the user side is not always perpendicular to the horizontal plane. The total number of ports on the fluid antenna is $M$, and the spacing between adjacent ports is $d_{\rm U}$. $d_{\rm BS}$ and $d_{\rm U}$ are both $\lambda/2$. Therefore, the total space occupied by the fluid antenna is $d_{\rm U}(M-1)$.  When receiving signals, $m_0$ ports are selected for signal reception. The indices of the selected ports are represented by the vector $\boldsymbol{r}=[r_1,\ldots,r_m,\ldots,r_{m_0}]^T\in\mathbb{Z}^{m_0\times1}$, where $r_m\in[1,M]$, $r_1< r_2< \ldots < r_{m_0}$. The coordinates of the activated port $r_m$ are $\left(0,k_{r_m}d_{\rm U}\right)$ where $k_{r_m}d_{\rm U}$ is given by
\begin{equation}\label{eq2}
k_{r_m}d_{\rm U}=\frac{2(r_m-1)-M+1}{2}d_{\rm U}.
\end{equation}

\subsection{Signal Transmission}
We extend the field response channel model from \cite{10318061} under near-field conditions. The transmit signal is denoted by $\boldsymbol{s} \in \mathbb{C}^{N\times1}\sim\mathcal{C}\mathcal{N}(\boldsymbol{0},\boldsymbol{Q})$ where $\boldsymbol{Q}\in\mathbb{C}^{N\times N}$ represents the transmit covariance matrix. The received signal is expressed as
\begin{equation}\label{eq3}
\boldsymbol{y}(\boldsymbol{r})=\boldsymbol{H}(\boldsymbol{r})\boldsymbol{s}+\boldsymbol{z},
\end{equation}
where $\boldsymbol{H}(\boldsymbol{r})\in \mathbb{C}^{m_0\times N}$ is the channel matrix from the transmit antennas at the BS to the activated ports in the fluid antenna and $\boldsymbol{z}\in\mathbb{C}^{m_0\times 1}\sim \mathcal{C}\mathcal{N}(\boldsymbol{0},\sigma^2\boldsymbol{\mathrm{I}}_m)$ is the complex additive white Gaussian noise with variance $\sigma^2$ for each element in $\boldsymbol z$.

In the far field, the channel response matrix only depends on the distance between the scatterers and the transmit/receive antennas. However, under near-field conditions, the channel response matrix also depends on the departure angles (AoDs) from the scatterers to the transmit antennas and the arrival angles (AoAs) from the scatterers to the received ports.

The number of transmit and receive paths are respectively denoted by $L_t$ and $L_r$. At the BS side, the elevation
and azimuth AoDs of the $p$-th ($1\leq p\leq L_t$) transmit path are denoted by $\theta_t^p \in [-\frac{\pi}{2},\frac{\pi}{2}]$ and $\phi_t^p \in [0,2\pi]$. The distance from the scatterer to the origin $O_t$ is given by $l_t^p$. In the $p$-th transmit path, the propagation path difference between the position of the $n$-th transmit antenna and the origin $O_t$ is
\begin{equation}\label{eq4}
\rho_t^p(n)=\sqrt{{l_t^p}^2+\left(k_nd_{\rm BS}\right)^2-2l_t^p k_nd_{\rm BS}\cos\left({\frac{\pi}{2}-\theta_t^p}\right)}-l_t^p.
\end{equation}
To simplify (\ref{eq4}), we use the approximation $\sqrt{1+x}\approx 1+\frac{1}{2}x-\frac{1}{8}x^2$ for small $x$ as the size of the antennas is generally much smaller than the distance from the scatterers to the center of the antennas. Hence, the simplified $\rho_t^p(n)$ can be represented by
\begin{equation}\label{eq5}
\rho_t^p(n)=-k_nd_{\rm BS} \sin{\theta_t^p}-\frac{k_n^2d_{\rm BS}^2\sin{\theta_t^p}^2}{2l_t^p}.
\end{equation}
Using the ratio of the propagation path difference $\rho_t^p(n)$ to the carrier wavelength $\lambda$, we can obtain the signal phase difference between the position of the $n$-th transmit antenna and origin $O_t$ in the $p$-th transmit path as $2\pi\rho_t^p(n)/\lambda$. Thus, the transmit field response vector can be written as 
\begin{equation}\label{eq6}
\boldsymbol{a}(n)\triangleq\left[e^{j\frac{2\pi}{\lambda}\rho_t^1(n)}, \ldots, e^{j\frac{2\pi}{\lambda}\rho_t^{L_t}(n)}\right]^T\in \mathbb{C}^{L_t\times1}.
\end{equation}
The field response vectors of all the $N$ transmit antennas can be expressed in matrix form as
\begin{equation}\label{eq7}
\boldsymbol{A}\triangleq[\boldsymbol{a}(1), \boldsymbol{a}(2), \ldots, \boldsymbol{a}(N)]\in\mathbb{C}^{L_t\times N}.
\end{equation}

Then considering the receiving field response matrix, we have the elevation and azimuth AoAs of the $q$-th ($1\leq q\leq L_t$) receive path, denoted by $\theta_r^q \in [-\frac{\pi}{2},\frac{\pi}{2}]$ and $\phi_r^q \in [0,2\pi]$. The distance from the scatterers to the origin $O_r$ in the $q$-th path is given by $l_r^q$. Similarly, in the $q$-th receive path, the propagation path difference between the position of the $r_m$-th activated port and the origin $O_r$ is expressed as
\begin{multline}\label{eq8}
\rho_r^q(r_m)=\\
\sqrt{{l_r^q}^2+(k_{r_m}d_{\rm U})^2-2 l_r^q k_{r_m}d_{\rm U}\cos\left({\frac{\pi}{2}-\theta_r^q}\right)}-l_r^q.
\end{multline}
As a consequence, we can simplify \eqref{eq8} using the approximation $\sqrt{1+x}\approxeq 1+\frac{1}{2}x-\frac{1}{8}x^2$ for small $x$ as
\begin{equation}\label{eq9}
\rho_r^q(r_m)=-k_{r_m}d_{\rm U}\sin{\theta_r^q}-\frac{k_{r_m}^2d_{\rm U}^2\sin{\theta_r^q}^2}{2l_r^q}.
\end{equation}
Again, the signal phase difference between the position of the $r_m$-th activated port and origin $O_r$ in the $q$-th receive path can be obtained by $2\pi\rho_r^q(r_m)/\lambda$. Hence, the receive field response vector can be written as 
\begin{equation}\label{eq10}
\boldsymbol{b}(r_m)\triangleq\left[e^{j\frac{2\pi}{\lambda}\rho_r^1(r_m)},\ldots, e^{j\frac{2\pi}{\lambda}\rho_r^{L_r}(r_m)}\right]^T\in \mathbb{C}^{L_r\times1}.
\end{equation}
The field response matrix of all $m_0$ activated ports is given by
\begin{equation}\label{eq11}
\boldsymbol{B}(\boldsymbol{r})\triangleq[\boldsymbol{b}(r_1),  \ldots, \boldsymbol{b}(r_{m_0})]\in\mathbb{C}^{L_r\times m_0}.
\end{equation}

Now, we define the path response matrix from the transmit antenna $O_t$ to the origin of the fluid antenna $O_r$ as $\boldsymbol{O}\in\mathbb{C}^{L_r\times L_t}$, with $O_{q,p}$ in the $p$-th row and $q$-th column referring to the response coefficient between the $p$-th transmit path and the $q$-th receive path. We assume that $O_{q,p}$ is independently and identically distributed (i.i.d.) Gaussian random variable with zero mean and variance $\alpha^2$. Overall, the end-to-end channel matrix, $\boldsymbol{H}(\boldsymbol{r})$, from the BS antennas to the activated ports of FAS at positions $\boldsymbol{r}$ can be written as 
\begin{equation}\label{eq12}
\boldsymbol{H}(\boldsymbol{r})=\boldsymbol{B}^H(\boldsymbol{r})\boldsymbol{O}\boldsymbol{A}\in \mathbb{C}^{m_0\times N}.
\end{equation}

Taking the expectation of the channel response matrix $\boldsymbol{O}$ over a period of time, we can obtain the achievable transmission rate as      
\begin{equation}\label{eq13}
R=\mathbb{E}_{\boldsymbol{O}}\left\{\log \det\left(\boldsymbol{\mathrm{I}}_{m_0}+\frac{1}{\sigma^2}\boldsymbol{H}(\boldsymbol{r})\boldsymbol{Q}\boldsymbol{H}^H(\boldsymbol{r})\right)\right\}.
\end{equation}

Considering the power consumption $\mathrm{tr}(\boldsymbol{Q})$ of the BS transmitting signals and the average static power consumption $P_c$ of a single user when the BS is operating, the energy efficiency $\eta(\boldsymbol{Q},\boldsymbol{r})$ of the communication system can be defined as
\begin{equation}\label{eq14}
\eta(\boldsymbol{Q},\boldsymbol{r})=\frac{\mathbb{E}_{\boldsymbol{O}}\left\{\log \det\left(\boldsymbol{\mathrm{I}}_{m_0}+\frac{1}{\sigma^2}\boldsymbol{H}(\boldsymbol{r})\boldsymbol{Q}\boldsymbol{H}^H(\boldsymbol{r})\right)\right\}}{\mathrm{tr}(\boldsymbol{Q})+P_c}.
\end{equation}
\subsection{Problem Formulation}
Our goal is to maximize the energy efficiency $\eta(\boldsymbol{Q},\boldsymbol{r})$ by optimizing $\boldsymbol{Q}$ and $\boldsymbol{r}$ jointly. Mathematically, the optimization problem can be formulated as 
\begin{subequations}\label{eq15}
\begin{align}
\max_{\boldsymbol{Q},\boldsymbol{r}} \quad & \frac{\mathbb{E}_{\boldsymbol{O}}\left\{\log \det\left(\boldsymbol{\mathrm{I}}_{m_0}+\frac{1}{\sigma^2}\boldsymbol{H}(\boldsymbol{r})\boldsymbol{Q}\boldsymbol{H}^H(\boldsymbol{r})\right)\right\}}{\mathrm{tr}(\boldsymbol{Q})+P_c}\tag{\ref{eq15}})\\
\textrm{s.t.} \quad & \boldsymbol{r}=\left[r_1,\ldots,r_m,\ldots,r_{m_0}\right]^T\in\mathbb{Z}^{m_0\times1},\label{eq:15a}\\
&r_m\in[1,M],\label{eq:15b}\\
&r_1< r_2< \cdots < r_{m_0},\label{eq:15c}\\
&\mathrm{tr}(\boldsymbol{Q})\leq P_{\rm max},\label{eq:15d}\\
&\boldsymbol{Q}\succeq \boldsymbol{0},\label{eq:15e}
\end{align}
\end{subequations}
where $P_{\rm max}$ is the maximum transmission power of the BS. Constraints (\ref{eq:15a}), (\ref{eq:15b}) and (\ref{eq:15c}) indicate that the numbers of activated ports are positive integers from $1$ to $M$, arranged in an ascending order. Constraint (\ref{eq:15d}) ensures that the power consumption does not exceed $P_{\rm max}$. Problem (\ref{eq15}) is difficult to solve as it involves a non-convex objective function with complex expressions and the variables $\boldsymbol{Q}$ and $\boldsymbol{r}$ are dependent. 
\section{Algorithm Design}
In this section, we first use the Jensen's inequality to obtain an upper bound on the transmission rate, and then perform alternating optimization on $\boldsymbol{Q}$ and $\boldsymbol{r}$ considering real-time constraints and limited resources. 
\subsection{Upper Bound of the Rate}
Using Monte Carlo methods to evaluate (\ref{eq13}) is complex and computationally expensive, we thus resort to simplifying the expression by deriving an upper bound for the transmission rate using Jensen's inequality \cite{Jensenineq}, yielding
\begin{align}
R&\leq\overline{R}\notag\\
&\triangleq \log \det \left( \boldsymbol{\mathrm{I}}_{m_0}+\frac{1}{\sigma^2}\mathbb{E}_{\boldsymbol{O}}\left\{\boldsymbol{H}(\boldsymbol{r})\boldsymbol{Q}\boldsymbol{H}^H(\boldsymbol{r})\right\}\right)\notag\\
&=\log\det \left(\boldsymbol{\mathrm{I}}_{m_0}+\frac{1}{\sigma^2}\boldsymbol{B}^H(\boldsymbol{r})\mathbb{E}_{\boldsymbol{O}}\left\{\boldsymbol{O}\boldsymbol{A}\boldsymbol{Q}\boldsymbol{A}^H\boldsymbol{O}^H\right\}\boldsymbol{B}(\boldsymbol{r})\right).\label{eq16}
\end{align}

Utilizing the statistical properties of matrices, the expectation in \eqref{eq16} can be expressed as \cite{Jensenineq}
\begin{equation}\label{eq17}
\mathbb{E}_{\boldsymbol{O}}\left\{\boldsymbol{O}\boldsymbol{A}\boldsymbol{Q}\boldsymbol{A}^H\boldsymbol{O}^H\right\}=\mathrm{tr}\left(\boldsymbol{A}\boldsymbol{Q}\boldsymbol{A}^H\right)\alpha^2\boldsymbol{\mathrm{I}}_{L_r}.
\end{equation}
Thus, we obtain a simple upper bound for $R$ as
\begin{equation}\label{eq18}
\overline{R}=\log\det \left(\boldsymbol{\mathrm{I}}_{m_0}+\frac{\alpha^2}{\sigma^2}\mathrm{tr}\left(\boldsymbol{A}\boldsymbol{Q}\boldsymbol{A}^H\right)\boldsymbol{B}^H(\boldsymbol{r})\boldsymbol{B}(\boldsymbol{r})\right).
\end{equation}

Though the upper bound of the rate, $R$, can avoid complex expectation calculations, Problem \eqref{eq15} remains a non-convex optimization problem. To tackle this, we adopt the method of alternating optimization to solve it.
\subsection{Transmit Covariance Matrix Optimization}
We first consider that the fluid antenna locations are fixed and focus on optimizing the transmission covariance matrix $\boldsymbol{Q}$ only. In what follows, we have the subproblem
\begin{subequations}\label{eq19}
\begin{align}
\max_{\boldsymbol{Q}} \quad & \frac{\log\det \left(\boldsymbol{\mathrm{I}}_{m_0}+\frac{\alpha^2}{\sigma^2}\mathrm{tr}\left(\boldsymbol{A}\boldsymbol{Q}\boldsymbol{A}^H\right)\boldsymbol{B}^H(\boldsymbol{r})\boldsymbol{B}(\boldsymbol{r})\right)}{\mathrm{tr}(\boldsymbol{Q})+Pc}\tag{\ref{eq19}}\\
\textrm{s.t.} \quad  &\mathrm{tr}(\boldsymbol{Q})\leq P_{\max},\\
&\boldsymbol{Q}\succeq \boldsymbol{0}.
\end{align}
\end{subequations}

To start with, we denote the numerator of the objective function \eqref{eq19} as $f(\boldsymbol{Q})$ and the denominator as $g(\boldsymbol{Q})$, with $g(\boldsymbol{Q})>0$. To solve \eqref{eq19}, we transform the optimization problem from a fractional form into a linear combination form \cite{6157574} with the balancing factor $\eta$ as shown below:
\begin{subequations}\label{eq20}
\begin{align}
\max_{\boldsymbol{Q}} \quad & f(\boldsymbol{Q})-\eta g(\boldsymbol{Q})\tag{\ref{eq20}}\\
\textrm{s.t.} \quad  &\mathrm{tr}(\boldsymbol{Q})\leq P_{\max},\\
&\boldsymbol{Q}\succeq \boldsymbol{0}.
\end{align}
\end{subequations}
We first select a value within the range of $\boldsymbol{Q}$ as the initial value for $\boldsymbol{Q}^{(0)}$. By calculating $\frac{f(\boldsymbol{Q}^{(0)})}{g(\boldsymbol{Q}^{(0)})}$, we can obtain the initial value of $\eta$, denoted by $\eta^{(0)}$. Substituting $\eta^{(0)}$ into the optimization problem \eqref{eq20}, we can obtain $\boldsymbol{Q}^{(1)}$. We can calculate $\frac{f(\boldsymbol{Q}^{(1)})}{g(\boldsymbol{Q}^{(1)})}$ as the iterative value of $\eta$, denoted by $\eta^{(1)}$. Continuing this iterative process until the value of optimization problem \eqref{eq20} falls within a very small interval $\left[-\epsilon,\epsilon\right]$ ($\epsilon>0,\epsilon\rightarrow 0$). 

The algorithm is described as Algorithm \ref{alg1}. 

\begin{algorithm}[H]
    \caption{Optimizing Transmit Covariance Matrix}\label{alg1}
    \begin{algorithmic}[1]
    \REQUIRE $M, N, m_0, L_t, L_r, \theta_t^p, \theta_r^q, \sigma^2, l_t^p, l_r^q, \alpha^2, \boldsymbol{O}, P_{\rm max},
    \newline \boldsymbol{r}, d_{\rm U},d_{\rm BS}, \lambda, P_c$.
    \ENSURE $\boldsymbol{Q}$
    \STATE Initialize $\boldsymbol{Q}^{(0)}$,threshold $\epsilon$, and calculate objective function $\eta(\boldsymbol{Q})$. Set iteration index $i=0$.
     \REPEAT
            \STATE $\eta^{(i)}=\frac{f(\boldsymbol{Q}^{(i)})}{g(\boldsymbol{Q}^{(i)})}$.
            \STATE Calculate $\boldsymbol{Q}^{(i+1)}=\arg \max_{\boldsymbol{Q}} f(\boldsymbol{Q})-\eta^{(i)} g(\boldsymbol{Q})$.
            \STATE Set $i=i+1$.
    \UNTIL{$|f(\boldsymbol{Q}^{(i+1)})-\eta^{(i)} g(\boldsymbol{Q}^{(i+1)})|\leq \epsilon$}.
    \STATE \textbf{Output}: Transmit Covariance Matrix $\boldsymbol{Q}^{(i+1)}$, $\eta^{(i)}$.
\end{algorithmic}
\end{algorithm}
\subsection{Fluid Antenna Position Optimization}
When the transmit covariance matrix $\boldsymbol{Q}$ is fixed, optimizing the positions of the receiving fluid antenna ports can be viewed as an optimization problem to maximize the transmission rate. So the port activation strategy is to maximize the total transmission rate of all ports. This subproblem can be formulated as
\begin{subequations}\label{eq21}
\begin{align}
\max_{\boldsymbol{r}} \quad & \log\det \left(\boldsymbol{\mathrm{I}}_{m_0}+\frac{\alpha^2}{\sigma^2}\mathrm{tr}\left(\boldsymbol{A}\boldsymbol{Q}\boldsymbol{A}^H\right)\boldsymbol{B}^H(\boldsymbol{r})\boldsymbol{B}(\boldsymbol{r})\right)\tag{\ref{eq21}}\\
\textrm{s.t.} \quad  &\boldsymbol{r}=\left[r_1,\ldots,r_m,\ldots,r_{m_0}\right]^T\in\mathbb{Z}^{m_0\times1},\\
&r_m\in[1,M],\\
&r_1< r_2< \cdots < r_{m_0}.
\end{align}
\end{subequations}

We approach this by first setting $\beta=\frac{\alpha^2}{\sigma^2}{\rm tr}(\boldsymbol{A}\boldsymbol{Q}\boldsymbol{A}^H)$. Then an upper bound on the achievable rate $\overline{R}$ can be derived as
\begin{align}\label{eq22}
\overline{R} & =\log\det \left(\boldsymbol{\mathrm{I}}_{m_0}+\beta\boldsymbol{B}^H(\boldsymbol{r})\boldsymbol{B}(\boldsymbol{r})\right),\nonumber\\
    &=\log\det \left(\boldsymbol{\mathrm{I}}_{L_r}+\beta\boldsymbol{B}(\boldsymbol{r})\boldsymbol{B}^H(\boldsymbol{r})\right),\nonumber\\
    &=\log\det \left(\boldsymbol{\mathrm{I}}_{L_r}+\beta\sum_{m=1}^{m_0}\boldsymbol{b}(r_m)\boldsymbol{b}^H(r_m)\right).
\end{align}
Using \eqref{eq22}, we can optimize one of the activated ports. To do so, by removing the $r_m$-th column from $\boldsymbol{B}$, the remaining $m_0-1$ columns are combined to form a new matrix
\begin{equation}\label{eq23}
\overline{\boldsymbol{B}}_m=[\boldsymbol{b}(r_1),\dots,\boldsymbol{b}(r_{m-1}),\boldsymbol{b}(r_{m+1}),\dots,\boldsymbol{b}(r_{m_0})].
\end{equation}
Then separating $\boldsymbol{B}$ from $\overline{\boldsymbol{B}}_m$ and $\boldsymbol{b}(r_m)$, the objective function in (\ref{eq21}) can be rewritten as
\begin{multline}\label{eq24}
\overline{R}=\log \det\left(1+\beta\boldsymbol{b}^H(r_m)\left(\boldsymbol{I}_{L_r}+\beta\overline{\boldsymbol{B}}_m\overline{\boldsymbol{B}}_m^H\right)^{-1}\boldsymbol{b}(r_m)\right)\\
+\log \det\left(\boldsymbol{I}_{L_r}+\beta\overline{\boldsymbol{B}}_m\overline{\boldsymbol{B}}_m^H\right).
\end{multline}
Therefore, maximizing the objective function $\overline{R}$ with all port positions unchanged except for the $r_m$-th activated port is equivalent to maximizing 
\begin{equation}\label{eq25}
p(r_m)=\boldsymbol{b}^H(r_m)\left(\boldsymbol{I}_{L_r}+\beta\overline{\boldsymbol{B}}_m\overline{\boldsymbol{B}}_m^H\right)^{-1}\boldsymbol{b}(r_m).
\end{equation}
Consequently, the optimization subproblem for the activated port $r_m$ can be formulated as
\begin{subequations}\label{eq26}
\begin{align}
\max_{r_m} \quad & \boldsymbol{b}^H(r_m)\left(\boldsymbol{I}_{L_r}+\beta\overline{\boldsymbol{B}}_m\overline{\boldsymbol{B}}_m^H\right)^{-1}\boldsymbol{b}(r_m)\tag{\ref{eq26}}\\
\textrm{s.t.} \quad  &\boldsymbol{r}=\left[r_1,\ldots,r_m,\ldots,r_{m_0}\right]^T\in\mathbb{Z}^{m_0\times1},\\
&r_m\in[1,M],\\
&r_1< r_2< \cdots < r_{m_0}.
\end{align}
\end{subequations}

When the total number of ports $M$ and the number of activated ports $m_0$ are not too large, we can iterate through all the combinations of activated port indices and use an exhaustive search to find the best combination of activated port indices that maximizes (\ref{eq26}). The overall proposed alternating algorithm is thus summarized in Algorithm~\ref{alg2}.

\begin{algorithm}[H]
    \caption{Transmission Rate Maximization}\label{alg2}
    \begin{algorithmic}[1]
    \REQUIRE $M, N, m_0, L_t, L_r, \theta_t^p, \theta_r^q, l_t^p, l_r^q, \sigma^2, \alpha^2, \boldsymbol{O}, P_{\rm max},
    \newline d_{\rm U},d_{\rm BS},\lambda$.
    \ENSURE $\boldsymbol{r},\boldsymbol{Q}$.
    \STATE Initialize $\boldsymbol{Q}^{(0)},\boldsymbol{r}^{(0)}$,threshold $\epsilon$, and calculate objective function $\eta^{(0)}(\boldsymbol{Q},\boldsymbol{r})$. Set iteration index $i=0$.
    \REPEAT
        \STATE Update $\boldsymbol{Q}$ by optimizing subproblem(\ref{eq20}).
        \FOR{$m=1$ to $m_0$}
            \STATE Calculate $\overline{\boldsymbol{B}}_m$ as (\ref{eq23}).
            \STATE Update $r_m$ by optimizing subproblem (\ref{eq26}).
        \ENDFOR
        \STATE $i=i+1$.
        \STATE Calculate $\eta^{(i)}(\boldsymbol{Q},\boldsymbol{r})$.
    \UNTIL $|\eta^{(i)}(\boldsymbol{Q},\boldsymbol{r})-\eta^{(i-1)}(\boldsymbol{Q},\boldsymbol{r})|\leq \epsilon$.
\end{algorithmic}
\end{algorithm}

\section{Simulation Results}
In this section, we demonstrate the performance of the proposed joint BS beamforming and FAS position optimization algorithm in near-field communications through computer simulations. In these simulations, the number of BS antennas $N$ is set to $20$ and the total number of ports for the fluid antenna at the user $M$ is $21$. The antenna spacing $d_{\rm BS}$ and port spacing $d_{\rm U}$ are both assumed to be $\lambda/2$. The number of transmit and receive paths for the channel is set to $L_t=L_r=3$. The AoDs $\theta_t^p$ and AoAs $\theta_r^q$ are both uniformly randomly distributed within $[-\pi/2,\pi/2]$. The carrier wavelength $\lambda$ is assumed to be $5~{\rm mm}$. The variance of $O_{q,p}$ is assumed to be $\alpha^2=1/L_r$. The average noise power is set as $\sigma^2=10~{\rm dBm}$. Also, the average static power consumption $P_c$ is set to $0.1~{\rm W}$. The signal-to-noise ratio (SNR) is defined as $P_{\rm max}/\sigma^2$. 

We will refer to Algorithm~\ref{alg2} as $\textbf{Proposed-FAS}$ and compare it with the following benchmark schemes:
\begin{itemize}
\item $\textbf{Random-FAS}$: The activated ports at the user's fluid antenna are randomly selected.
\item $\textbf{Conventional FPA (fixed-position antenna)}$: In this case, both the BS and the user are equipped with conventional fixed-position antennas, with the user having two fixed-position antennas for reception, spaced apart by $d_{\rm U}(M-1)$. 
\end{itemize}

\begin{figure}[t]
	\centering
	\begin{minipage}{0.49\linewidth}
		\centering
		\includegraphics[width=0.9\linewidth]{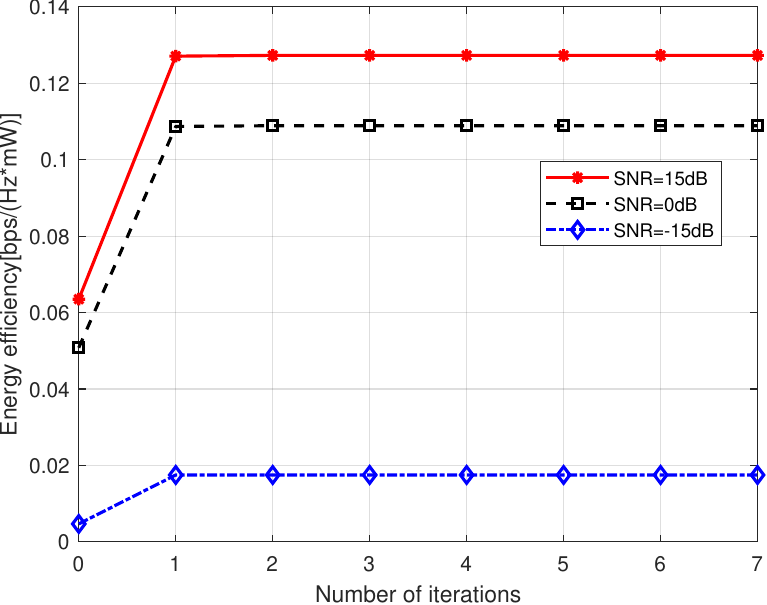}
		\caption{Convergence behaviour of Algorithm \ref{alg2} at different SNR.}
		\label{fig2}
	\end{minipage}
	\begin{minipage}{0.49\linewidth}
		\centering
		\includegraphics[width=0.9\linewidth]{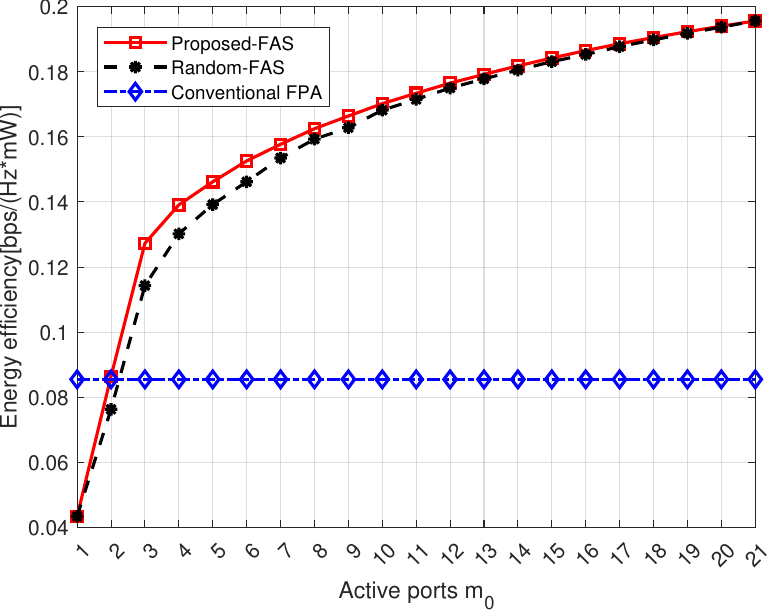}
		\caption{Energy efficiency against the number of activated ports.}
		\label{fig3}
	\end{minipage}
\end{figure}

In Fig.~\ref{fig2}, we show the convergence performance of Algorithm~\ref{alg2} under different values of SNR or $P_{\rm max}$. It is shown that Algorithm~\ref{alg2} converges fast and the energy efficiency is already very close to the final value after the first iteration. Due to the high complexity of the exhaustive algorithm with many fluid antenna ports, Algorithm~\ref{alg2} uses alternating optimization to achieve a local maximum that closely approximates the optimal solution.

\begin{figure}[t]
	\centering
	\begin{minipage}{0.49\linewidth}
		\centering
		\includegraphics[width=0.9\linewidth]{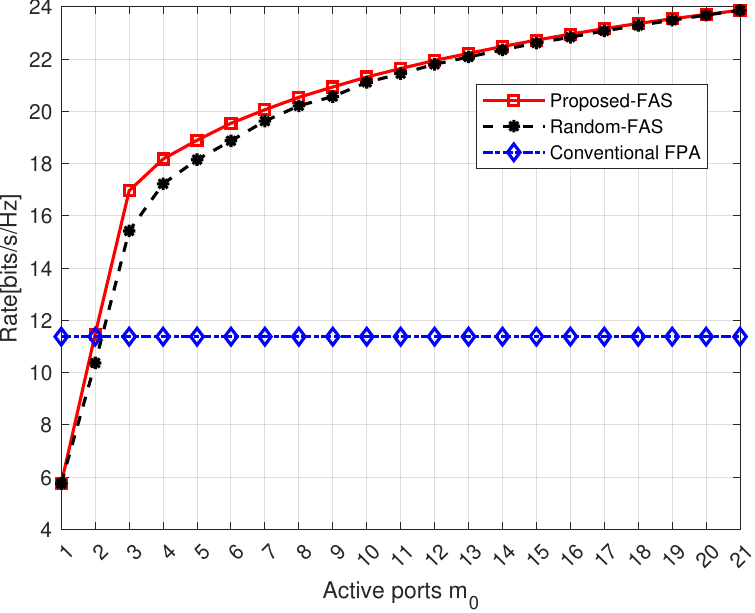}
		\caption{Achievable rate against the number of activated ports.}
		\label{fig5}
	\end{minipage}
	\begin{minipage}{0.49\linewidth}
		\centering
		\includegraphics[width=0.9\linewidth]{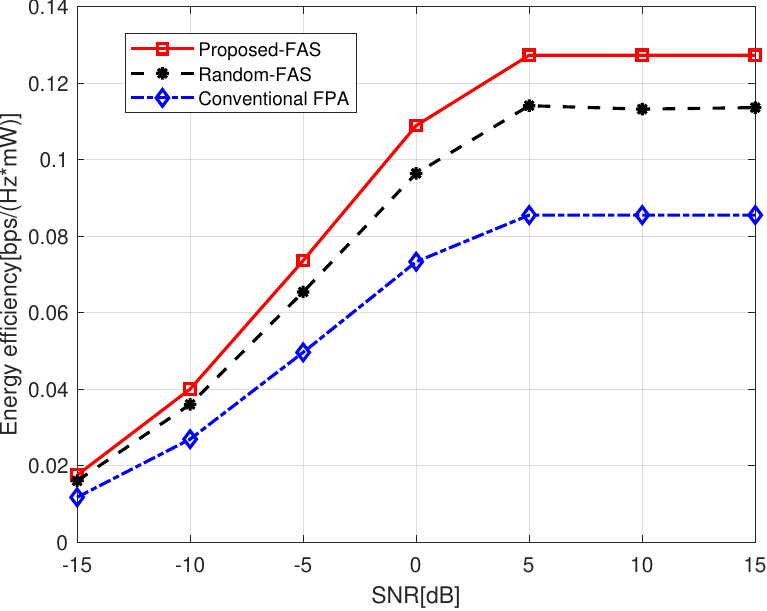}
		\caption{Performance comparison against different SNR.}
		\label{fig4}
	\end{minipage}
\end{figure}

With ${\rm SNR} = 15~{\rm dB}$, Fig.~\ref{fig3} and Fig.~\ref{fig5} compare the energy efficiency and achievable rate of three different designs under varying numbers of activated ports. In Fig.~\ref{fig3} and Fig.~\ref{fig5}, the performance of `Conventional FPA' remains unchanged while `Proposed-FAS' and `Random-FAS' exhibit increasing performance as the number of activated ports increases, with the rate of growth gradually slowing down. Specifically, when $m_0 = 3$, `Proposed-FAS' obtains $6.76\%$ and $62.69\%$ energy efficiency gains and $9.97\%$ and $49.07\%$ rate gains compared to `Random-FAS' and `Conventional FPA'. 
when the number of activated ports grows, `Random-FAS' gradually approaches `Proposed-FAS' performance due to the the higher chance of selecting better positions. 

Fig.~\ref{fig4} illustrates the variation in energy efficiency of three different designs under different values of SNR. The number of activated ports $m_0$ is set to $3$ in this figure. When the SNR is around $5~{\rm dB}$ or lower, the energy efficiency of the three designs increases as the SNR increases. In contrast, when the SNR is greater than $5~{\rm dB}$, the energy efficiency no longer increases, which means that we have achieved the maximum energy efficiency within the constraints of limited SNR. 

\section{Conclusion}
This letter proposed an algorithm to jointly optimize the BS transmit beamforming and the FAS activated ports at the user in the near field for maximizing the energy efficiency. In this letter, we formulated the problem and constructed an alternating optimization procedure. Simulation results demonstrated that compared to baseline algorithms, the proposed algorithm can achieve considerably higher energy efficiency. In the future, the relevant optimization ideas and algorithms can be applied to near-field communication in two-dimensional (2D) FAS.

\bibliographystyle{IEEEtran}

\end{document}